\documentclass[aip,rsi,reprint]{revtex4-2}

\usepackage{graphicx}
\usepackage{dcolumn}
\usepackage{bm}

\usepackage{xcolor}

\usepackage[utf8]{inputenc}
\usepackage[T1]{fontenc}
\usepackage{mathptmx}

\usepackage[colorlinks = true,
linkcolor = blue,
urlcolor  = blue,
citecolor = blue,
anchorcolor = blue,
breaklinks]{hyperref}

\usepackage{float}
\usepackage[per-mode=symbol]{siunitx}
\DeclareSIUnit{\sqrthz}{\ensuremath{\text{\hertz}^{-1/2}}}
\DeclareSIUnit{\sqrtf}{\ensuremath{\text{f}^{-1/2}}}
\DeclareSIUnit{\decibelm}{\ensuremath{\text{dBm}}}

\begin{document}
\title[]{Wideband current modulation of diode lasers for frequency stabilization}

\author{Tilman Preuschoff}
 \email{apqpub@physik.tu-darmstadt.de}
\author{Patrick Baus}
\affiliation{Technische Universit\"at Darmstadt, Institut f\"ur Angewandte Physik, Schlossgartenstra\ss e 7, 64289 Darmstadt, Germany}
\author{Malte Schlosser}
\affiliation{Technische Universit\"at Darmstadt, Institut f\"ur Angewandte Physik, Schlossgartenstra\ss e 7, 64289 Darmstadt, Germany}
\author{Gerhard Birkl}
\homepage{http://www.iap.tu-darmstadt.de/apq}
\affiliation{Technische Universit\"at Darmstadt, Institut f\"ur Angewandte Physik, Schlossgartenstra\ss e 7, 64289 Darmstadt, Germany}
\affiliation{Helmholtz Forschungsakademie Hessen für FAIR (HFHF), Campus Darmstadt, Schlossgartenstra\ss e 2, 64289 Darmstadt, Germany}

\date{\today}

\begin{abstract}
We present a current modulation technique for diode laser systems that is specifically designed for high-bandwidth laser frequency stabilization and wideband frequency modulation with a flat transfer function. It consists of a dedicated current source and an impedance matching circuit both placed close to the laser diode. The transfer behaviour of the system is analysed under realistic conditions employing an external cavity diode laser (ECDL) system. We achieve a phase lag smaller than \SI{90}{\degree} up to \SI{25}{\mega\hertz} and a gain flatness of \SI{\pm3}{\decibel} in the frequency range of DC to \SI{100}{\mega\hertz} while the passive stability of the laser system is not impaired. The potential of the current modulation scheme is demonstrated in an optical phase-locked loop between two ECDL systems with phase noise of \SI{42}{\milli\radian}${}_{\text{rms}}$. The design files are available as an open-source project.
\end{abstract}

\maketitle

\section{Introduction}

Diode laser systems with narrow linewidth and wideband frequency-modulation capabilities play an essential role in many experiments in quantum optics and photonics. Sideband generation for various techniques such as Pound-Drever-Hall stabilization \cite{Black2001} or frequency-modulation spectroscopy \cite{Demtroeder2015} benefits from a wideband laser current modulation with a well defined transfer behaviour. Active frequency stabilization with a large bandwidth of the feedback loop is required to achieve a sufficient linewidth reduction since diode lasers inherently show high-frequency noise. Techniques such as Pound-Drever-Hall stabilization to a high-finesse optical cavity \cite{Drever1983,Black2001} or optical phase-locked loops (OPLL)\cite{Santarelli1994, Appel_2009} offer feedback bandwidths up to several tens of MHz. Utilizing the full potential of these techniques requires a wideband servo input acting on the laser frequency with a flat response function and a low phase lag.
 
A large frequency-modulation bandwidth can be achieved by applying a control voltage directly to the laser diode using a bias tee.\cite{Yim2014} This servo input exhibits the nonlinear characteristics of the laser diode since the forward voltage is modulated instead of the forward current. Additionally, it reduces the noise immunity of the system and is prone to damaging of the laser diode. A commonly used improved method relies on a junction-gate field-effect transistor (JFET) in parallel to the laser diode. \cite{Appel_2009} The JFET is used to bypass the laser diode driving current. The current splitting ratio is given by the impedance of the laser diode. Thus, the current-modulation transfer function of the circuit is strongly dependent on the laser diode operating conditions, such as forward current and temperature. A linear response is only achieved if the transistor operating point is tuned depending on the laser diode voltage. Also, the passive stability of the system is degraded by an offset current of up to several mA through the JFET that is subject to thermal drifts. This can be avoided by using a precision voltage-controlled bipolar current source that injects a modulated current into the laser diode. Typically, this is implemented via a modulation input at the diode-laser current driver as suggested by Libbrecht and Hall. \cite{Libbrecht1993} However, the bandwidth of this implementation is limited by the impedance mismatch between current source and laser diode as well as by delays induced by the cable connecting the current driver and  the laser system. \cite{Erickson2008}

In order to avoid these drawbacks, we developed a simple technique for laser frequency modulation comprising a wideband, flat modulation response with a low phase lag independent of the laser diode impedance, a linear response over the entire operating range, and a negligible influence on the passive stability of the system. 
This is achieved by integrating a fast current source 
complemented by a passive impedance matching network close to the laser diode on the printed circuit board of the laser backplane. Design details of the backplane are available as an open-source hardware project. \cite{APQGit_LaserHead} The transfer function of the circuit is characterized using an external cavity diode laser (ECDL) system. The quality of the impedance matching is demonstrated by comparison to a system without impedance matching. Furthermore, the transfer behaviour in the time domain is studied by recording a large-signal step response.  In order to illustrate the potential of the presented method, the circuit is used as the servo input of an OPLL between two ECDL systems.
\begin{figure}[htb]
	\includegraphics[width = 0.48 \textwidth]{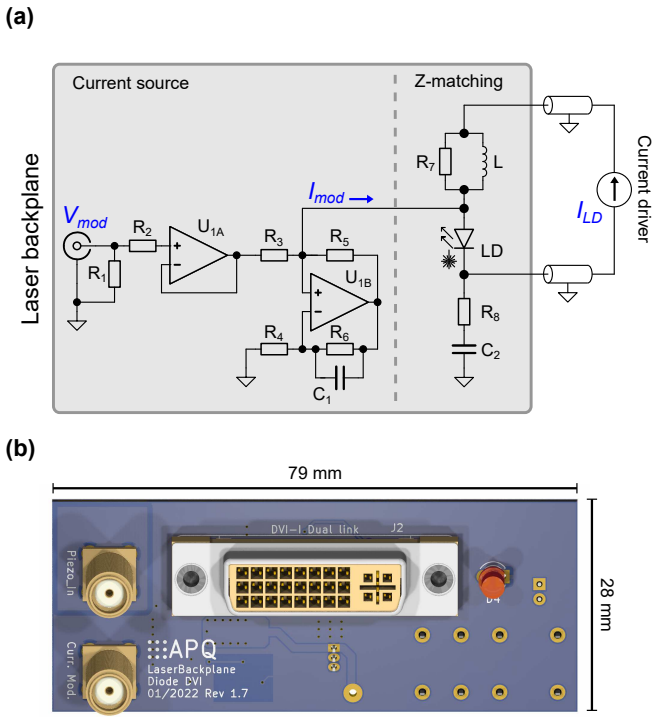} 
	\caption{
		\label{fig:circuit} (a) Simplified circuit schematics of the current source and the impedance matching network (Z-matching). Dual operational amplifiers $U_{1A}$, $U_{1B}$: Analog Devices ADA4807-2. Resistors: $R_1, R_7 = \SI{49.9}{\ohm}$, $R_2$ to $R_6 = \SI{1}{\kilo\ohm}$, $R_8 = \SI{4.7}{\ohm}$. Capacitors: $C_1 = \SI{1.5}{\pico\farad}$, $C_2 = \SI{10}{\nano\farad}$. Inductor: $L = \SI{33}{\micro\henry}$. Laser diode ${LD}$: Thorlabs L785H1.
		(b) Rendered 3D-model of the printed circuit board. 
	}
\end{figure}%
\section{Circuit design}
The current source used in this work is a \textit{Howland Current Pump}. \cite{Sheingold1964} 
As depicted in Fig.~\ref{fig:circuit}(a) it is formed by a high speed operational amplifier $U_{1B}$ (Analog Devices ADA4807-2) that features a high DC precision and a rail-to-rail output. The circuit injects a current $I_{mod}$ proportional to the input voltage $V_{mod}$ directly into the laser diode. The transconductance gain of the current source is set by the resistors $R_3$ to $R_6 = $\,\SI{1}{\kilo\ohm} to $I_{mod}/V_{mod} = \SI{1}{\milli\ampere\per\volt}$. In order to obtain a high output impedance, the resistors $R_3$ to $R_6$ are matched to \SI{\pm0.05}{\percent} using a precision resistor network (Vishay MORNTA1001A).
The absolute accuracy of the resistor network yields an accuracy of \SI{\pm0.1}{\percent} for the transconductance gain. A small capacitance $C_1$ in the negative feedback loop of $U_{1B}$ guarantees stability of the circuit. A second amplifier $U_{1A}$ (Analog Devices ADA4807-2) is used as an input buffer, providing a \SI{50}{\ohm}-input ($R_1$). The input series-resistance ($R_2$) is set to form a low-pass filter with the input capacitance of the amplifier reducing the gain peaking and defining a proper roll-off for high frequencies.  The offset current induced by the current source can be estimated from the maximum input bias current and offset voltage of the the operational amplifier both stated in the data sheet \cite{ADA4807} to a value below \SI{3}{\micro\ampere} with a thermal drift below \SI{10}{\nano\ampere\per\kelvin}. The output current noise of the source as estimated from the operational amplifier voltage and current noise \cite{ADA4807} is well below \SI{30}{\pico\ampere\per\sqrt\hertz}.  The compliance voltage of the source $V_c$ is given by half of the positive output swing of the operational amplifier. The dual amplifier is supplied with \SI{\pm5}{\volt} by two small-outline voltage-regulators (Analog Devices LT1761-5 and Analog Devices LT1964-5) also integrated in the laser backplane (not shown) resulting in $V_{c} = \SI{2.4}{\volt}$ and a \SI{\pm4.8}{\volt} voltage range for $V_{mod}$ accounting for the required head-room of the input and output voltages of the operational amplifiers. The obtained compliance voltage $V_{c}$ is sufficient as forward voltage for most infrared laser diodes for which the circuit was configured. If necessary, $V_{c}$ can be doubled by operating $U_{1B}$ with a single-ended \SI{+10}{\volt} supply.

For most applications, the laser diode is supplied with a current $I_{\mathit{LD}}$ using a cable with a typical length of several meters. Reflections of the modulation signal unintentionally transmitted through this cable at the high-impedance output of the laser current driver induce resonances in the transfer function depending on the cable length and impedance. These resonances are suppressed by impedance matching with the resistor $R_7$. In order to reduce heat dissipation in the laser head and the required compliance voltage of the current driver, the DC load at $R_7$ is reduced employing a compact shielded power inductor $L$ (Coilcraft XGL5050). In order to maintain impedance matching for low frequencies, the value of $L$ is chosen as large as possible accounting for size constraints and the maximum driving current. In order to optimize the noise immunity of the laser system, a floating configuration for the laser diode is desirable, e.g.~in order to avoid ground loops. In this case, the modulation input ground reference is not connected to the laser diode cathode in the laser head.  Thus, additional reflections occur in the return path of the modulation current. This can be avoided by providing a low impedance node for high modulation frequencies with the capacitor $C_2$ while a high impedance is maintained at low noise-frequencies (snubber). A small series resistance $R_8$ guarantees stability of the laser current driver under capacitive load. The values of $C_2$ and $R_8$ are chosen empirically for best performance. Note, that a current modulation via JFET or bias tee always requires a common ground reference of the laser diode and the modulation input in the laser head thereby reducing the noise immunity of the system by design.

The circuit presented in Fig.~\ref{fig:circuit}(a) is implemented as a compact laser backplane depicted in Fig.~\ref{fig:circuit}(b) which has been made available as an open-source design. \cite{APQGit_LaserHead} The printed circuit board also contains a protection circuit for the laser diode consisting of a relay and a reverse-voltage protection diode (not shown). The backplane is designed to be connected to our custom ultra-low noise current driver \footnote{P. Baus \textit{et. al.}, unpublished} using a low-capacity digital visual interface (DVI) cable (Supra Cables HD5). The driver also provides the supply voltages for the current source. All measurements presented here have been conducted in this configuration, unless stated otherwise. In general, the modulation current source can also be used with standard current drivers in combination with an external power supply. Note, that the laser diode voltage is limited to $V_c$ in case the laser diode cathode is referenced to the modulation input ground. 
\begin{figure*}[htp]
	\includegraphics[width = \textwidth]{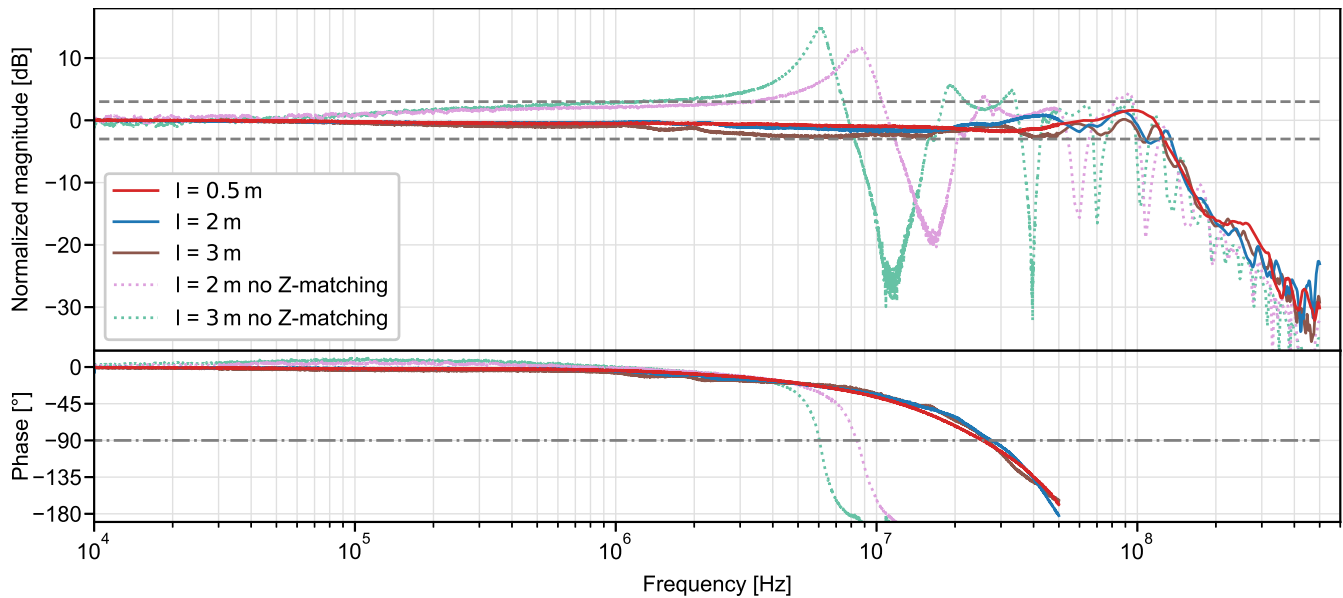} 
	\caption{
		\label{fig:transfer} Transfer function of the modulation input for different cable lengths $l$ with (solid lines) and without (dotted lines) impedance matching network (Z-matching). (top) The magnitude is normalized to the gain at \SI{10}{\kilo\hertz}. The dashed, grey lines indicate the \SI{\pm3}{\decibel}-band. (bottom) The phase lag between input and output has been measured for frequencies up to \SI{50}{\mega\hertz}. The dashed-dotted, grey line marks a phase lag of \SI{90}{\degree}. 
	}
\end{figure*}%
\section{Current modulation transfer function}\label{sec:transfer}
In this work, the wideband current source depicted in Fig.~\ref{fig:circuit} is applied to a custom ECDL system \cite{Baillard2006, Martin2016} emitting light at \SI{780}{nm}.  
For the modulation parameters studied in this work, the current modulation of the laser diode is proportional to the induced intensity modulation.\cite{Petermann1988} Hence, the transfer function of our current modulation input can be studied by analysing the laser intensity modulation resulting from a modulation of $V_{mod}$.
For investigation of the frequency modulation transfer function, the complex behaviour of the laser diode \cite{Kobayashi1982}  has to be taken into account which is application specific.

\begin{figure}[htb]
	\includegraphics[scale = 1]{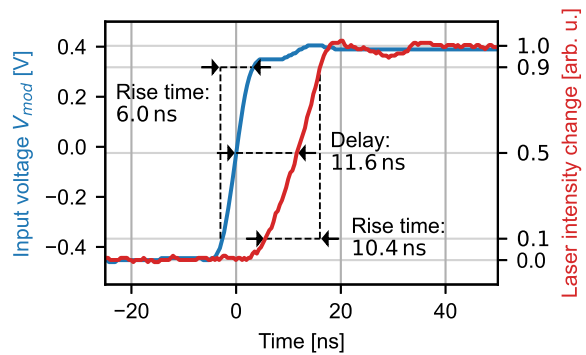} 
	\caption{
		\label{fig:step} Response of the laser system to a step in $V_{mod}$ with an amplitude of \SI{0.85}{\volt} (average over 64 repetitions). 
	}
\end{figure}

For modulation frequencies below \SI{50}{\mega\hertz}, we detect the resulting laser intensity modulation with a custom DC photodetector based on a transimpedance amplifier with a bandwidth of \SI{150}{\mega\hertz}. This detector allows for a measurement of the transfer-function phase. 
The transfer function is recorded using a Bode analyser (Omicron Lab Bode100). For the scalar measurement of the transfer function above \SI{50}{\mega\hertz}, a spectrum analyser with tracking generator (Rohde \& Schwarz FPC1500) in combination with a fast photodiode amplified by a \SI{500}{\mega\hertz}-bandwidth RF-amplifier is used. For both frequency ranges, the power level at the modulation input is set to \SI{-20}{\decibelm}.  Figure~\ref{fig:transfer} shows the transfer function for different lengths $l$ of the DVI cable used to supply the laser. As a reference, the impedance matching is omitted ('no Z-matching') by removing $C_2$ and setting $R_1 = \SI{0}{\ohm}$ (see Fig.~\ref{fig:circuit}(a)) for the cable lengths $l=\SI{2}{\meter}$ and $l=\SI{3}{\meter}$. For this setting, the system shows pronounced resonances with frequency and magnitude dependent on $l$ (dotted lines). If the impedance matching network  is set in place (solid lines), the resonances are strongly suppressed, such that  a gain flatness better than \SI{\pm3}{\decibel} is obtained for frequencies up to \SI{100}{\mega\hertz} and cable lengths up to \SI{3}{\meter}. The phase response of the system is depicted in Fig.~\ref{fig:transfer}~(bottom). The remaining influence of resonances is negligible. We obtain a phase lag less than \SI{90}{\degree} up to \SI{25}{\mega\hertz}  for all cable lengths. A strong increase of the phase lag is observable, if the impedance matching network is not included.\\
The dynamical behaviour of the system is studied in the time domain by recording the large-signal step response. For that purpose, the change in laser intensity is monitored with a fast photodetector (bandwidth \SI{500}{\mega\hertz}) and a fast oscilloscope when a rectangular signal with an amplitude of \SI{0.85}{\volt} and a rise time of \SI{6.0(5)}{\nano\second} is applied to the modulation input. The operating parameters of the laser system have been adjusted in such a way that the induced current change results in a mode-hop free frequency step.  Figure~\ref{fig:step} shows the input signal and the response of the system averaged over 64 repetitions. The measurement has been corrected for the delay induced by the cables in the signal paths of the measurement setup. We observe a delay of \SI{11.6(5)}{\nano\second} induced by the current source. Accounting for an increased rise time of \SI{10.4(5)}{\nano\second}, this result is in good agreement with a phase lag of \SI{90}{\degree} at \SI{25}{\mega\hertz} corresponding to a delay of \SI{10}{\nano\second}. The observable amount of ringing and overshoot is negligible for most applications.
\section{Application: Optical phase lock of two diode laser systems}
The presented wideband current source is designed as a servo input for high-bandwidth laser frequency stabilization. One possible application is an OPLL between two ECDL systems.\cite{Appel_2009} The laser system used for characterization of the current source (Sec.~\ref{sec:transfer}) is locked to an identical ECDL system with an offset frequency of \SI{3}{\giga\hertz}. For both laser systems, current drivers adapting the design suggested in Ref.~\cite{Libbrecht1993} are used. In particular, these drivers feature a standard current modulation input. In order to implement the OPLL, a beat note is detected using a fast photodetector. The beat signal is divided by four (Analog Devices HMC365) and compared to a reference signal provided by a synthesizer (Agilent E4432) using a digital phase frequency detector (Analog Devices HMC439). 
The full OPLL consists of two feedback stages acting on one of the ECDL systems: a frequency lock and a phase lock. 
In the first stage, an intermediate-bandwidth frequency lock is implemented using the digital proportional-integral controller based on the STEMlab platform and the locking scheme both described in our previous work. \cite{Preuschoff2020} The modulation input of the current driver is used as a servo input. The second stage (phase lock) defines the phase noise performance of the OPLL, thus requiring a large bandwidth. It is implemented by a proportional-derivative (PD) controller acting on the fast modulation input of the presented laser-backplane. For the PD controller, an optimal corner frequency of \SI{1}{\mega\hertz} is determined.

\begin{figure}[htb]
	\includegraphics[scale=0.99]{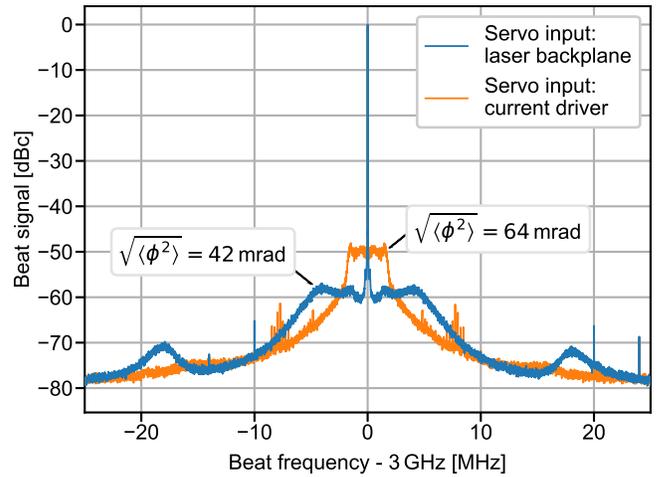} 
	\caption{
		\label{fig:phaselock} Beat note spectrum of two ECDL systems phase locked with an offset frequency of \SI{3}{\giga\hertz} using the frequency-modulation technique presented in this work (blue) and the modulation input included in the laser current driver for comparison (orange).  The resolution bandwidth of this measurement is \SI{10}{\kilo\hertz}. 
	}
\end{figure}

Figure~\ref{fig:phaselock} shows the spectrum of the beat note (blue) evaluated by a spectrum analyser (Tektronix RSA306) set to a resolution bandwidth of \SI{10}{\kilo\hertz}. For comparison, the input signals for the first and the second stage are swapped (orange) such that the performance of the OPLL is determined by the performance of the modulation input of the current driver. Using the laser-backplane current source in the fast servo branch, the bandwidth of the servo loop as indicated by the frequency of the noise-gain peaking is increased from \SI{1.5}{\mega\hertz} to \SI{4.5}{\mega\hertz}. The bandwidth achieved in this configuration is consistent with a total delay introduced by the feedback loop estimated to \SI{55}{\nano\second} resulting in a phase lag of approximately \SI{90}{\degree} at \SI{4.5}{\mega\hertz}. The rms phase noise $\sqrt{\bigl<\phi^2\bigr>}$ of the beat signals depicted in Fig.~\ref{fig:phaselock} is determined by integrating the normalized single-sideband power-spectral-density from \SI{100}{\kilo\hertz} to \SI{50}{\mega\hertz} \cite{Riehle2004} yielding a significant reduction of phase noise from \SI{64}{\milli\radian}${}_{\text{rms}}$ (orange) to \SI{42}{\milli\radian}${}_{\text{rms}}$ (blue). Note, that for offset frequencies below \SI{100}{\kilo\hertz}, the blue spectrum is dominated by the phase noise of the spectrum analyser. Since the impedance matching network is active for both measurements, the enhanced performance can be attributed to the superior performance of the fast laser-backplane current source. The measured bandwidth of the OPLL is not limited by the phase lag of the servo input. We expect that it can be increased further by reducing delays in the feedback loop.  

\section{Conclusion}
In this work, we present a current modulation technique that is utilized as a servo input for high-bandwidth laser frequency stabilization. Employing a fast current source guarantees a high performance at DC vastly independent of the laser diode operating and ambient conditions by design, i.e.~a constant transconductance gain of \SI{1}{\milli\ampere\per\volt} over the entire input voltage range and a thermally stable offset current below \SI{3}{\micro\ampere}. Additionally, the current source allows for a floating operation of the laser diode increasing the noise immunity of the setup.  For most laser systems, the passive stability is not degraded due to the low offset current and low current noise induced by the source. Hence, this modulation technique can also be used in open-loop applications such as fast frequency and intensity changes with arbitrary waveforms. The current source placed close to the laser diode in the laser head is complemented by a passive impedance matching network that effectively damps resonances occurring due to reflections in the laser-system current-supply cable. The presented network has only to be matched to the nominal impedance of the cable in use. Due to the typically low AC-impedance of laser diodes we expect this network to be suitable for a wide range of laser diodes without modifications. The open-source availability of the design of the laser backplane facilitates reproduction and adaption to custom applications, such as connectors, pin configurations, and power supplies. \\
We characterize the transfer function of the modulation input under realistic operating conditions using an ECDL system supplied by a DVI cable. A gain flatness of \SI{\pm3}{\decibel} up to \SI{100}{\mega\hertz} and a phase lag less than \SI{90}{\degree} up to \SI{25}{\mega\hertz} is achieved for cable lengths up to \SI{3}{\meter}. This measurement also allows us to record bandwidth-limiting resonances depending on the cable length if the impedance matching network is removed. Analysis of the large-signal step-response in the time domain yields a delay of \SI{11.6(5)}{\nano\second} induced by the current source in good agreement with the phase lag obtained from the transfer function.\\
Most laser frequency stabilization setups involve signal and light paths of several meters. In this case, the achievable bandwidth will not be limited by the current modulation technique presented here. When compared to other common current modulation techniques, our scheme combines the advantages of a direct current injection used in laser current drivers \cite{Libbrecht1993} with the high bandwidth offered by a modulation input employing a JFET or a bias tee. \cite{Appel_2009,Yim2014} We illustrate the benefits of our fast modulation input in a typical OPLL application. As compared to the standard modulation input of a current driver, the control bandwidth has been increased by a factor of three resulting in a significant reduction of the phase noise.
\section*{Supplementary Material}
The supplementary material contains the CAD design files for KiCAD 6, the bill of materials, and additional assembly instructions. Also provided are industry standard Gerber files directly accepted by printed circuit board manufacturers.
\section*{Acknowledgements}
We acknowledge financial support by the Deutsche Forschungsgemeinschaft (DFG) [Grant BI 647/6-2, DFG SPP 1929 GiRyd], by the Federal Ministry of Education and Research (BMBF) [Grants 05P21RDFA1 and 13N15981], HGS-HIRe, and HFHF.
\section*{Author Declarations}
\subsection*{Conflict of Interest}
	The authors have no conflicts to disclose.
\section*{Data availability}
	The data and circuit board design files that support the findings of this study are openly available in the GitHub repository ``TU-Darmstadt-APQ/Laser\_Backplane\_DVI'', reference number \cite{APQGit_LaserHead}.
\section*{References}
\bibliography{CurrentMod_arxiv}
\end{document}